\documentclass[aps,prb,twocolumn,superscriptaddress,showpacs,amsmath,amssymb]{revtex4-1}

\usepackage{graphicx}
\usepackage{bm}
\begin{document}

\title{Changing the type of superconductivity by magnetic and potential scattering}
\author{V. G. Kogan}
\email{kogan@ameslab.gov}
\affiliation{Ames Laboratory - DOE and Department of Physics,Iowa State University, Ames, IA 50011}

\author{R. Prozorov}
\email{prozorov@ameslab.gov}
\affiliation{Ames Laboratory - DOE and Department of Physics,Iowa State University, Ames, IA 50011}

\begin{abstract} By evaluating the upper and thermodynamic critical fields, $H_{c2}$ and $H_c$, and their ratio   $H_{c2}/H_c $  at arbitrary temperatures, we argue that  situations are possible when a type-II material is transformed to type-I  by adding magnetic impurities.
\end{abstract}
\date{16 October 2014 }
\maketitle

\section{Introduction}

Traditionally,  classification  of isotropic superconductors as type-I and II is based on the value of the Ginzburg-Landau parameter $\kappa = \lambda/\xi$,  the ratio of the London penetration depth and the coherence length. At the critical temperature $T_c$, $\kappa=1/\sqrt{2}$ separates type-II materials for which $\kappa>1/\sqrt{2}$ from the type-I where $\kappa<1/\sqrt{2}$. At $T_c$,  the upper and thermodynamic critical fields are related by $ H_{c2}=\sqrt{2}\kappa H_c$ so that their ratio $R=H_{c2}/H_c=1$ if $\kappa=1/\sqrt{2}$. In type-II materials $ H_{c2}>H_c$ that is equivalent to $\kappa>1/\sqrt{2}$, in other words, at $T_c$ the criteria for the type-II  $\kappa>1/\sqrt{2}$ and $R>1$ are equivalent. However, for  $T<T_c$, where the critical fields as well as $\lambda$
 and $\xi$ can be evaluated using the microscopic theory,\cite{Eil} the relation $ H_{c2}=\sqrt{2}\kappa H_c$ no longer holds. For the type-II superconductivity we need of course $H_{c2}>H_c$ or $R>1$. Hence, the GL $\kappa$ criterion cannot be used, except at $T_c$.\cite{KP-last}

Parameter $R$ was calculated in the clean limit for  anisotropic   Fermi surfaces and   order parameters and possibilities of changes in the type of superconductivity with temperature or the applied field direction were demonstrated.\cite{KP-last}
Still, the question arises of how  $R$ depends on scattering? It is well known that non-magnetic impurities cause $H_{c2}$ to increase.\cite{HW} Effect of magnetic impurities  is just the opposite,  $H_{c2}$ is suppressed.\cite{AG,KP} On the other hand, the thermodynamic critical field $H_c$  (along with the condensation energy) is insensitive to the transport scattering, but is reduced by the pair-breaking.\cite{AG,Skalsky,Maki} It is  difficult to evaluate  scattering effects for arbitrary anisotropies, here we limit ourselves to isotropic s-wave superconductors with arbitrary transport and magnetic  scattering  (in Born approximation).

Evaluating $H_{c2}$ we use extensively   Ref.\,\onlinecite{KP} to which we refer the readers interested in derivation of formulas we employ. For   $H_c$ we make use of the review by  Maki.\cite{Maki} We show that pair-breaking scattering may suppress  $H_{c2}$ faster than  $H_{c}$ and even cause transition from the type-II to the type-I superconductivity. By using specific material parameters  for Th,   we show a  way of making this, initially type-I superconductor, to become type-II upon introduction of non-magnetic disorder, and then turn it back  to type-I by \textit{adding} magnetic scattering. Our theory contains  a single material  parameter - the clean-limit   $\kappa_{GL}$ at $T_c$ - so that our results can be readily applied to  other isotropic materials.
\section{  $\bm {T\to T_c} $}

Here we compare the slopes of $ H_{c2}(T) $ and $H_c(T)$ at $T_c$.
We use  dimensionless variables
\begin{equation}
t = \frac{T}{T_c}\,,\qquad h= H_{c2}\frac{\hbar^2v^2}{2\pi T_c^2 \phi_0} \,,
\label{t,h}
\end{equation}
and the scattering parameters
\begin{equation}
 \rho  =
\frac{\hbar}{2\pi T_c\tau }\,,\quad \rho_m = \frac{\hbar}{2\pi T_c\tau_m}\,,\quad \rho^\pm =\rho\pm\rho_m\,,
\label{rhos}
\end{equation}
where $\tau$ and $\tau_m$ are   times of the  transport and pair-breaking  scattering. At $T_c$  one has for isotropic s-wave materials: \cite{KP}
\begin{eqnarray}
 &&    -\frac{dh}{dt}\Big|_{t=1}=3 \rho_- ^2\left[1-\rho_m\psi^\prime \left(
\rho_m +\frac{1}{2}\right)\right]  \Big/\qquad \qquad \qquad \nonumber\\
&&\left[ \psi\left(\rho_m +\frac{1}{2}\right)
- \psi\left(\frac{\rho^++1}{2}\right) + \frac{\rho_-}{2} \psi^\prime \left(
\rho_m +\frac{1}{2}\right)\right] ,\qquad\label{h'_GL}
\end{eqnarray}
where $\psi$ is the digamma function.
The actual slope is
\begin{equation}
 \frac{d H_{c2}}{dT}\Big|_{T_c} =  \frac{2\pi  \phi_0T_c}{\hbar^2v^2} \frac{dh}{dt}\Big|_{t=1}\,.
\label{t,h}
\end{equation}

According to Ref.\,\onlinecite{Maki} (or \onlinecite{Skalsky}):
\begin{eqnarray}
\frac{dH_{c}}{dT}\Big|_{T_c}= -2\pi \sqrt{8\pi N(0)}\,\frac{1-\rho_m\psi^\prime  (
\rho_m + 1/2) }  { \sqrt{b_1(\rho_m)}}\,,\nonumber\\
b_1=-\frac{1}{2} \psi^{\prime\prime} \left( \rho_m +\frac{1}{2}\right)- \frac{\rho_m}{6} \psi^{\prime\prime\prime} \left( \rho_m +\frac{1}{2}\right);
 \label{dHc/dT}
\end{eqnarray}
$N(0)$ is the density of states on the Fermi level per spin.
The ratio of our interest is:
\begin{eqnarray}
&&R(T_c)=\frac{ H_{c2} }{ H_{c} }\Big|_{T_c}= \frac{dH_{c2}/dT}{dH_{c}/dT}\Big|_{T_c} = \nonumber\\
&& C\frac{T_c}{T_{c0}}\,\frac{ \rho_-^2 \sqrt{-\frac{1}{2} \psi^{\prime\prime} \left( \rho_m +\frac{1}{2}\right)- \frac{\rho_m}{6} \psi^{\prime\prime\prime} \left( \rho_m +\frac{1}{2}\right)}} {\psi\left(\rho_m +\frac{1}{2}\right)
- \psi\left(\frac{\rho^++1}{2}\right) + \frac{\rho_-}{2} \psi^\prime \left(
\rho_m +\frac{1}{2}\right)}
\,, \qquad \label{Rc}\\
&&C=\frac{3\phi_0k_BT_{c0}}{ \hbar^2 v^2\sqrt{8\pi N(0)} } \,.
 \label{C}
\end{eqnarray}
The constant $C $ is, in fact, close to $\kappa_{GL}$ for the clean case:
\begin{eqnarray}
\kappa_{GL}=  \frac{3 \phi_0 T_{c0}  }{ \hbar^2v^2    \sqrt{7\zeta(3)\pi N(0)}} =C\sqrt{\frac{8}{7\zeta(3)}}=0.975\, C  \,.\qquad
  \label{kapGL}
\end{eqnarray}
According to Abrikosov-Gor'kov (AG),\cite{AG}
\begin{eqnarray}
 \frac{T_c}{  T_{c0}}= \exp\left[ -\psi  \left( \rho_m +\frac{1}{2}\right)+\psi  \left(  \frac{1}{2}\right)\right]
  \label{AG}
\end{eqnarray}
One can check that for $\rho_m\to 0$, $R(T_c)\to \sqrt{2}\,\kappa_{GL}$.

  \begin{figure}[t]
\begin{center}
\includegraphics[width=8.5cm] {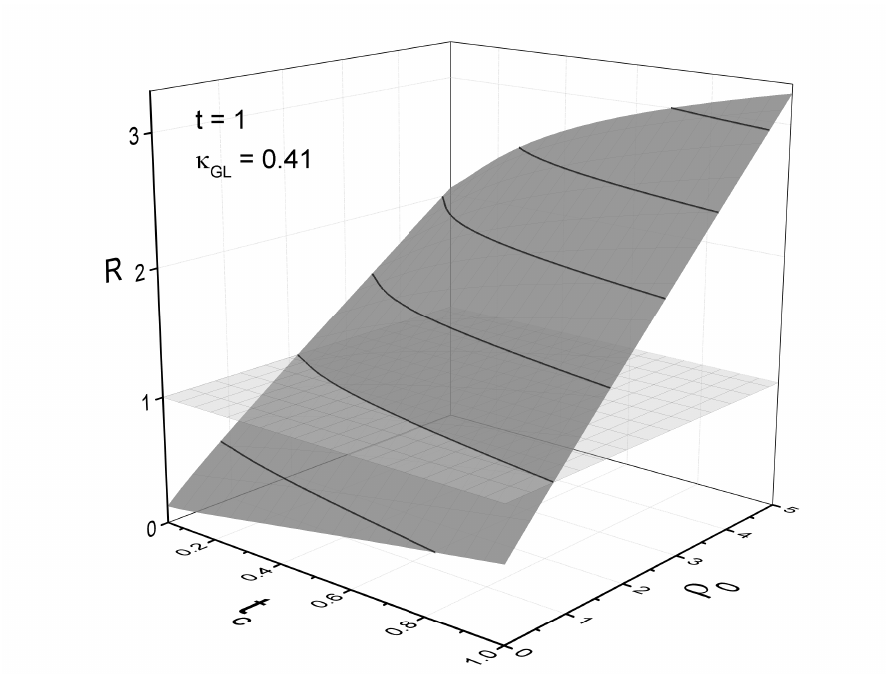}
\caption{  The ratio of slopes of $H_{c2}$ and $H_c$ at $T_c$ for $\kappa_{GL} =0.41$.   Instead  of  the second coordinate $0<\rho_{m0}<0.14$ we use directly $1>t_c=T_c/T_{co}>0$. The horizontal plane is $R=1$.
}
\label{f1}
\end{center}
\end{figure}

The scattering parameters (\ref{rhos})  are  convenient in analytical  work. Being   $T_c$ dependent, while $T_c$ depends on $\rho_m$, they are not good for comparison with data. It is better to work with
 $T_c$ independent material parameters
\begin{equation}
\rho_{m0} = \frac{\hbar}{2\pi T_{c0}\tau_m}=\rho_m \frac{T_c}{T_{c0}},\,\,\, \rho_0  =
\frac{\hbar}{2\pi T_{c0}\tau }=\rho \frac{ T_c}{T_{c0}}\,.
\label{rhos0}
\end{equation}
  It is  better yet  to plot the ratio $R$  as a function of $\rho_0$ and $t_c=T_c/T_{c0}$ instead of $\rho_{m0}$. The way to do this is described in  Appendix A. Results of numerical evaluation of $R(T_c)$ are shown in Fig.\,\ref{f1}. We have chosen for $\kappa_{GL} =0.41$ having in mind possible applications for Th.\cite{Doug}
The ratio $R$ increases nearly linearly with increasing transport $\rho_0$ for any $t_c$. However,  for a fixed $\rho_0$, $R$ decreases when the pair-breaking intensifies.  Of course, for large enough transport scattering (in this case for $\rho_0>5$)
the material is type-II for any pair-breaking. We show below that for relatively clean samples   unusual scenarios are possible.

\section{$\bm{T=0}$}

\begin{figure}[t]
\begin{center}
\includegraphics[width=8.cm] {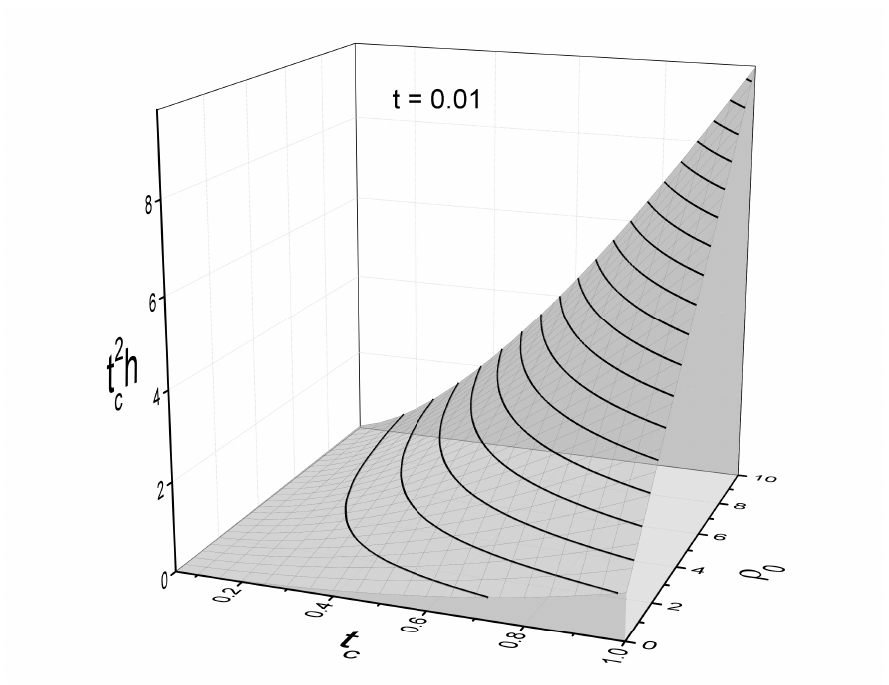}
\caption{ The quantity  $t_c^2h =H_{c2}(0)/H^*$   is plotted versus    $ \rho_{0} $ and   $t_c=T_c/T_{co} $.  To reduce the calculation time  $t=0.01$ is taken instead of $t=0$, }
\label{f2}
\end{center}
\end{figure}

The dimensional quantity $h$ of Eq.\,(\ref{t,h})
for $t\to 0$ has been evaluated in Ref.\,\onlinecite{KP} for any $\rho$ and $\rho_m$. The upper critical field is given by
\begin{equation}
H_{c2} = \frac{2\pi T_{c0}^2 \phi_0}{\hbar^2v^2}\, t_c^2h= H^* t_c^2h\,.
\label{Hc20}
\end{equation}
The constant $H^*$ is  close to the clean limit  $H_{c2}(0)= \pi T_{c0}^2 \phi_0e^{2-\gamma}/2\hbar^2v^2$ ($\gamma=0.577$ is the Euler constant);\cite{HW}  in fact, $H^*/H_{c2}(0)=4 e^{\gamma-2}=0.964$.
The dimensionless $t_c^2h$ versus $\rho_0$ and $t_c $ is shown in Fig.\,\ref{f2}.

 We now calculate $H_c(0)=\sqrt{8\pi F(0)}$, where  evaluation of
the condensation energy $F(0)$ requires knowledge of the order parameter $\Delta(\rho_m)$. The latter satisfies: \cite{AG}
\begin{eqnarray}
 \frac{\Delta}{ \Delta_0}&=& e^{-  \pi\zeta/4 }   \,,\quad \zeta=\frac{\hbar}{\tau_m\Delta}\leq 1\,, \label{z<1}\\
 \frac{\Delta}{ \Delta_0}&=& e^{  \eta(\zeta)
} \,,\quad \zeta > 1\,,
 \label{z>1}\\
\eta&=& -\cosh^{-1}\zeta- \frac{\zeta}{2} \sin^{-1}\frac{1}{\zeta}+\frac{\sqrt{ \zeta^2-1}}{2\zeta},\qquad
  \label{eta}
 \end{eqnarray}
where $\Delta_0=T_{c0}\pi e^{-\gamma}$ is the zero-$T$ gap of a clean sample.

The condensation energy at $T=0$  is given by:\cite{Skalsky,Maki}
\begin{eqnarray}
 \frac{2F(0)}{N(0)\Delta^2}&=& 1- \frac{\pi} {2 } \zeta + \frac{2} {3} \zeta^2  \,,\quad \zeta \leq 1\,, \label{17a}\\
 \frac{2F(0)}{N(0)\Delta^2}&=&1-   \zeta \sin^{-1}  \frac{1} { \zeta} + \zeta^2\left(1-\sqrt{1-1/\zeta^2}\right)\nonumber\\
&-& \frac{\zeta^2}{3}\left(1-(1-1/\zeta^2 )^{3/2}\right)\,,\quad \zeta > 1 .
 \label{17b}
\end{eqnarray}
It is convenient to normalize $F$ on the clean limit value
\begin{eqnarray}
F_0(0)= \frac{N(0)\Delta_0^2}{2}=\frac{N(0)T_{c0}^2 \pi^2}{2e^{2\gamma}}\,.
\label{F00}
  \end{eqnarray}
Then, Eq.\,(\ref{17a}) for $\zeta \leq 1$ transforms to:
\begin{eqnarray}
  f=\frac{F}{F_0}=e^{-\pi\zeta/2} \left(1- \frac{\pi} {2 } \zeta + \frac{2} {3} \zeta^2 \right) \,,
  \label{19}
  \end{eqnarray}
where Eq.\,(\ref{z<1}) has been used.  Clearly, $f=1$ for $\zeta=0$.

For $\zeta > 1$, we have:
\begin{eqnarray}
  f&=&\frac{F}{F_0}=e^{2\eta} \Big[1-   \zeta \sin^{-1}  \frac{1} { \zeta} + \zeta^2\left(1-\sqrt{1-1/\zeta^2}\right)\nonumber\\
&-&\frac{\zeta^2}{3}\left(1-(1-1/\zeta^2 )^{3/2}\right) \Big] \,. \label{eq20}
  \end{eqnarray}
We obtain:
\begin{eqnarray}
H_c =\sqrt{8\pi F_0 f}=\Delta_0\sqrt{4\pi  N(0)f(\rho_m) }=H_{c0}\sqrt{ f(\rho_m)},\qquad
 \label{Hc0}
  \end{eqnarray}
where $H_{c0}$ is the thermodynamic critical field at $T=0$ in the absence of magnetic impurities.

\begin{figure}[h]
 \begin{center}
 \includegraphics[width=10cm]{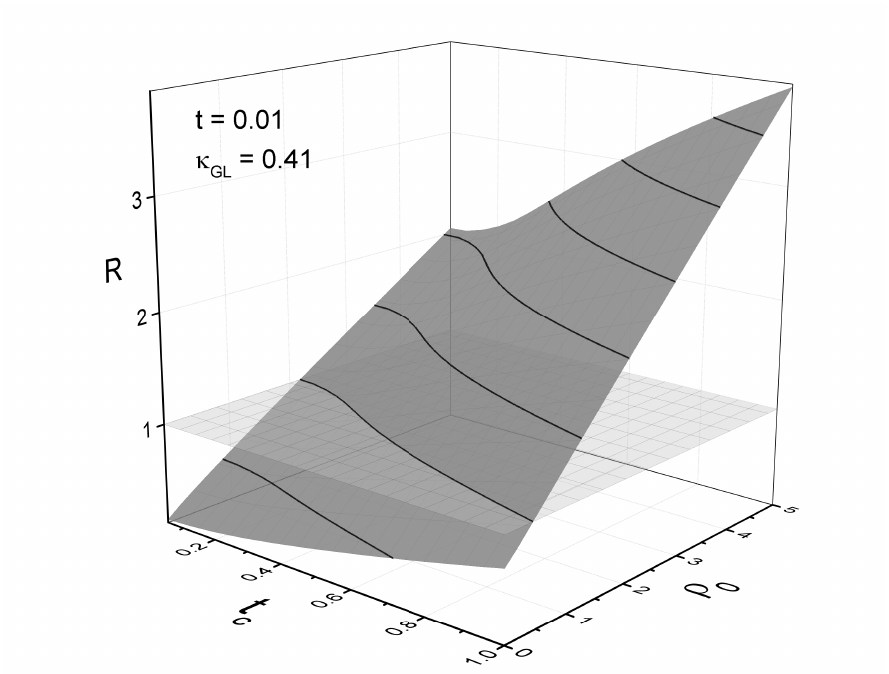}
 \caption{ The   ratio $R(0)=H_{c2}/H_c$ at $T=0$   for $ \kappa_{GL}=0.41$.   The   plane  $R=1$, the boundary between two types of superconductivity, is also shown.
  }
  \end{center}
  \label{f3}
 \end{figure}

Now   $R(0)=H_{c2}/H_c$ at $T=0$  can be evaluated:
\begin{eqnarray}
R(0) = \frac{\phi_0T_{c0}e^{\gamma}}{\hbar^2v^2\sqrt{ \pi  N(0)} }\frac{t_c^2h}{\sqrt{ f }}
=D \frac{t_c^2h}{\sqrt{ f }} \,,
 \label{R0}
  \end{eqnarray}
where the dimensionless constant is given by
 \begin{eqnarray}
D = \frac{\phi_0T_{c0}e^{\gamma}}{\hbar^2v^2\sqrt{ \pi  N(0)} } =\frac{e^\gamma \sqrt{  7\zeta(3) }}{3}\,\kappa_{GL} \approx 1.72\, \kappa_{GL}\,.\qquad
 \label{C0}
  \end{eqnarray}
 The ratio  $R(0)$  evaluated numerically is shown in Fig.\,\ref{f3} for $ \kappa_{GL}=0.41$.

\begin{figure}[htb]
 \includegraphics[width=8cm]{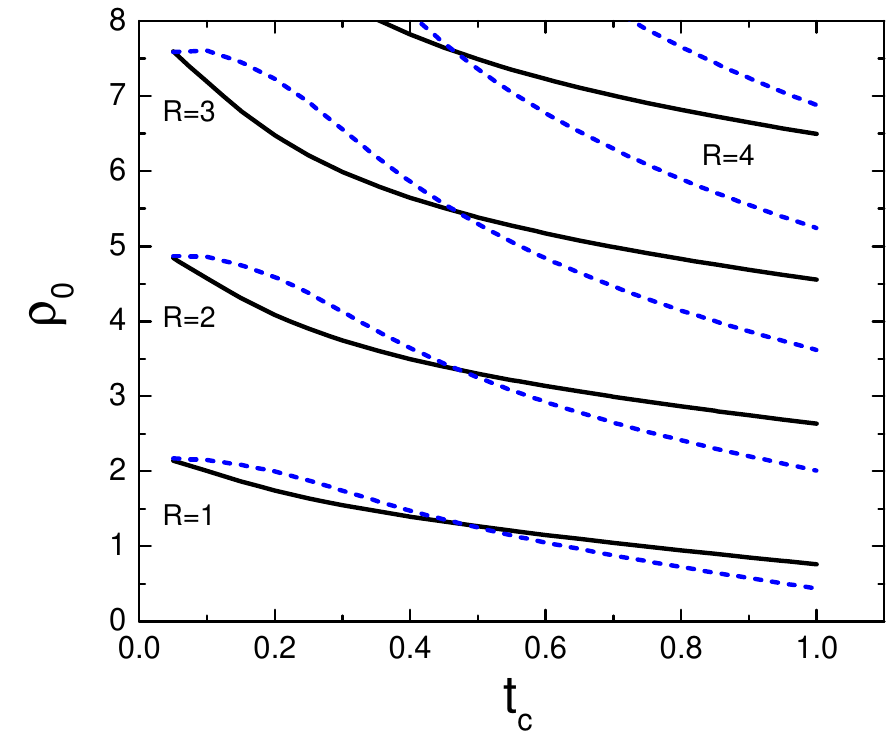}
 \caption{(Color online)   Solid lines: contours of $R(t_c,\rho_0)=$ const   for $T=T_c$. Dashed lines are for  $T=0$.  Both sets are calculated for $ \kappa_{GL}=0.41$.  }
  \label{f4}
 \end{figure}

Figure \,\ref{f4} shows curves along which $R(\rho_0,t_c)=$ const at $T=T_c$ and  $T=0$. The contour  $R(\rho_0,t_c)=1$ separates the upper part of the plane $ (t_c,\rho_0) $ where $R>1$ and which  corresponds to the type-II and the lower part where $R<1$, the type-I.
We note that the curves of constant $R(T_c)$ and $R(T=0)$ meet in the limit of strong pair-breaking; this can be shown  analytically, but we omit a cumbersome proof.  This result  could be expected:   when  $t_c\to 0$, the temperature  domain between 0 and $T_c$ shrinks to zero as well and, therefore, we must have $R(0)=R(T_c)$.

 Also, we observe that these curves cross at a certain level of pair-breaking, e.g., at  $t_c^*\approx 0.5$ for $R=1$. Clearly, the point $t_c^* $ separates narrow domains between solid and dashed lines on the phase plane $ (t_c,\rho_0) $ with different behavior of the material  in question with changing  temperature.
  Within the narrow part where  $t_c>t_c^* $, the material is type-I near $T_c$ and becomes type-II at some temperature on cooling toward 0. Within the narrow region where  $t_c<t_c^* $, it is of the type-I near $0$ and becomes type-II at some temperature on warming toward $T_c$.

\section{Discussion}

In many new-discovered materials, superconductivity coexists or competes with antiferromagnetism. Fe-based materials are an example. Pair-breaking scattering is likely to be present and  affect superconducting properties in various ways. Our work for isotropic s-wave case does not pretend to adequately describe these properties, given that the only material parameter entering our theory is  $\kappa_{GL}$. Although oversimplified, the isotropic example  may provide  a qualitative guidance to such questions as the type of superconductivity, needed for understanding the material behavior in magnetic fields.

Besides, these questions have a fundamental relevance since historically the classification of superconductors to two types came from the GL theory which holds only near the critical temperature. It is long known that most of pure elemental metals  are of the type-I and can be turned type-II by addition of non-magnetic scatterers.   We  show here that the magnetic scattering may reduce $H_{c2}$ faster than $H_c$ thus pushing materials toward type-I.

\begin{figure}[htb]
 \includegraphics[width=8cm]{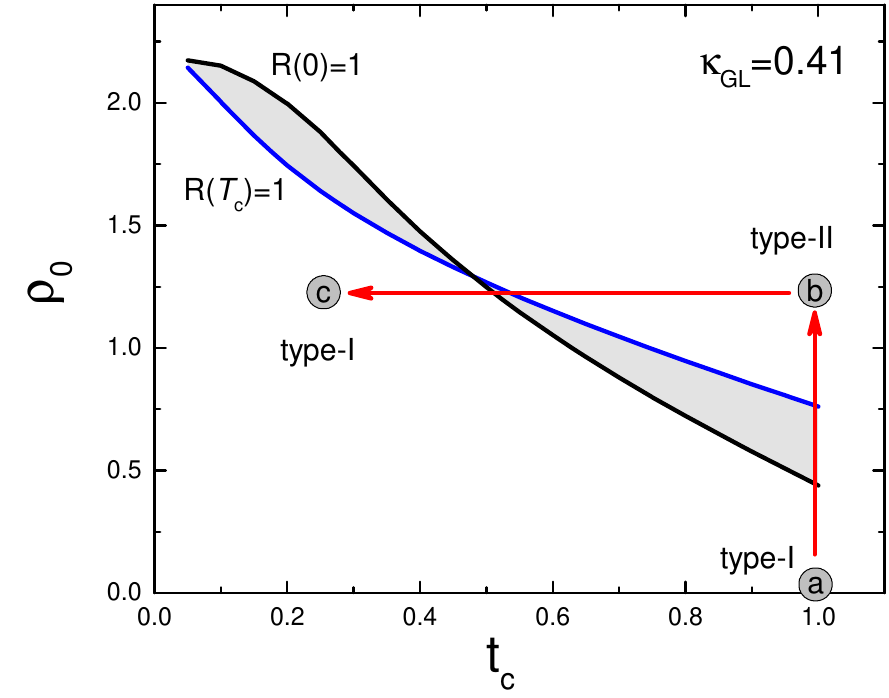}
 \caption{(Color online)   Contours of $R(t_c,\rho_0)=1$   for $T=T_c$ and  $T=0$ for $ \kappa_{GL}=0.41$.  }
  \label{f5}
 \end{figure}

One of such scenarios is illustrated in Fig.\,\ref{f5} for a material with $\kappa_{GL}=0.41$ (the value reported for the clean Th). Let the originally type-I clean material (point a) be doped with non-magnetic impurities to become  type-II (point b). If now pair-breaking impurities are added, $t_c$ is reduced and the material may reach the region under contours $R(0; t_c,\rho_0)=1$ and $R(T_c; t_c,\rho_0)=1$ which corresponds to type-I at all $T$s. We note that the type-I superconductivity emerging due to magnetic impurities is likely to be gapless, since it may appear at small $t_c$.

Interesting questions about possible change of the superconductivity type with changing temperature (for   material parameters in shaded areas of Fig.\,\ref{f5}), possible fluctuations near  this transition, peculiarities of intervortex interactions ect, are out of the scope of this article, and may be addressed if a material with proper characteristics is found. \\

The authors are grateful to  D. Finnemore,  J. Kirtley,   J. Thompson, P. Canfield, S. Bud'ko, V. Taufor, and B. Maple  for many helpful discussions.  The Ames Laboratory is supported by the Department of Energy, Office of  Basic Energy Sciences, Division of Materials Sciences and Engineering under Contract No. DE-AC02-07CH11358.

\appendix

\section{Numerical procedure}

The scattering parameters $\rho $ and $\rho_m$ of Eq.\,(\ref{rhos}) are not convenient variables for comparison with experimental data; they depend on
 $T_c$ which itself is determined by the Eq.\,(\ref{AG}) containing $\rho_m(T_c)$.  Instead one can use material parameters $\rho_0 $ (roughly, the ratio of the BCS coherence length and the mean free path) and $\rho_{m0}$  proportional to the pair-breaking scattering rate. Furthermore,   $\rho_{m0}$ is uniquely related by Eq.\,(\ref{AG}) to the actually measured $T_c$, so that $\rho_0$ and $T_c$ are the two independent parameters we use to present our results.

 Within our numerical scheme, this amounts to: (1) for a given $t_c$, solving Eq.\,(\ref{AG}) for $\rho_m$; then (2) forming $\rho_m$ and $\rho$ according to Eq.\,(\ref{rhos0}) in terms of $\rho_{m 0}$ and $\rho_0$, and (3), having $\rho_m$ and $\rho$, evaluating the slopes of $H_{c2}$ and $H_c$ at $T_c$, or calculating numerically $H_{c2}(0)$ and $H_c(0)$ with the help of the procedure developed in Ref.\,\onlinecite{KP-last} where Mathlab has been used.

\section{Strong pair-breaking}

Given a quite involved numerical procedure for  evaluation of the ratios $R(t_c,\rho_0)$, it is useful to have analytic results at least at some limiting points. We consider here the strong pair-breaking limit $t_c\to 0$. In this case, $h(0)$, related to   $H_{c2}(0)$, has been derived in Ref.\,\onlinecite{KP-last}:
 \begin{equation}
h(0)= \frac{1}{8}\left(\frac{\rho^-
}{\rho_m }\right)^2\left(\frac{\rho^-}{2\rho_m}+\ln
\frac{2\rho_m}{ \rho^+}\right)^{-1}  \,.
\label{h_gapless}
\end{equation}

Since all $\rho\propto 1/T_c$, their ratios and  $h(0)$ are $T_c$ independent. In particular,  \begin{equation}
\frac{\rho}{\rho_m}  = \frac{\rho_0}{\rho_{m0}}= 4e^\gamma\rho_0
\label{r/rm}
\end{equation}
where we replaced $\rho_{m0}$ with its critical value $\rho_{m0,cr}=e^{-\gamma}/4$. Hence, in this limit $h(0)$ depends only on $\rho_0$.

The critical fields at $T=0$ are
\begin{equation}
H_{c2} = \frac{2\pi T_{c}^2\phi_0}{\hbar^2v^2 }\,h(0)\,,\quad H_c=\frac{\Delta^2\tau_m\sqrt{\pi N(0)}}{\hbar\sqrt{3} }\,,
\label{critHs}
\end{equation}
were $H_c(0)$ is obtained by going to $\zeta\to\infty$ in the free energy (\ref{17b}). Taking into account that at $T=0$ for the strong pair-breaking $\Delta^2=2\pi^2 T_c^2$,\cite{AG,Maki} we obtain:
 \begin{equation}
R(0)|_{t_c\to 0}= \frac{\sqrt{3}\phi_0}{\pi \hbar v^2\sqrt{\pi N(0)}} \frac{h(0;\rho_0,\rho_{m0,cr})}{\tau_m}\Big|_{t_c\to 0}\,.
\label{R(0)tc=0}
\end{equation}
Here, $\tau_m\to 2\hbar/\Delta_0$, the critical value at which $t_c=0$.   This can also be written in terms of $\kappa_{GL}$ of Eq.\,(\ref{kapGL}):
 \begin{equation}
R(0)|_{t_c\to 0}= \frac{7\zeta( 3)e^{-\gamma}} {2\sqrt{3}} \kappa_{GL}h(0;\rho_0,\rho_{m0,cr}) \,.
\label{R(0)tc=0}
\end{equation}
By setting here $R(0)=1$ and solving numerically this equation for a given $\kappa_{GL} $, we can find the value of $\rho_0$ at which the curve $R(0; t_c,\rho_0)=1$ reaches $t_c=0$. This gives $\rho_0=2.21$ for $\kappa_{GL}=0.41$ in agreement with what is shown in Fig.\,\ref{f5}.
Similarly, we can verify all other limiting points of both  $R(0; t_c,\rho_0)=1$ and $R(T_c; t_c,\rho_0)=1$ to confirm our numerical results.

           \end{document}